\documentclass[twocolumn]{aastex7}

\usepackage{natbib}
\usepackage{amsmath}
\usepackage{url}
\usepackage{longtable}
\usepackage{graphicx}
\usepackage{multirow}
\usepackage{color}
\graphicspath{{paperfigures/}}

\newcommand{\err}[2]{\ensuremath{^{+#1}_{-#2}}}

\newcommand\ddfrac[2]{\frac{\displaystyle #1}{\displaystyle #2}}
\def\lsim{\hbox{\rlap{\raise 0.425ex\hbox{$<$}}\lower 0.65ex\hbox{$\sim$}}}
\def\gsim{\hbox{\rlap{\raise 0.425ex\hbox{$>$}}\lower 0.65ex\hbox{$\sim$}}}

\def\arcsec{\hbox{$^{\prime\prime}$}}
\def\arcdeg{\mbox{$^\circ$}}

\newcommand{\halpha}{H$\alpha$}
\newcommand{\hbeta}{H$\beta$}
\newcommand{\hgamma}{H$\gamma$}
\newcommand{\hdelta}{H$\delta$}

\shorttitle{Binary Neutron Star Delay-Time Distribution}
\shortauthors{Foley}

\begin{document}

\title{Missing Mergers: On the Unexpected Delay-Time Distribution of Binary Neutron Star Mergers}

\author[0000-0002-0786-7307]{Ryan~J.~Foley}
\affiliation{Department of Astronomy and Astrophysics, University of California, Santa Cruz, CA 95064, USA}
\email{foley@ucsc.edu}

\begin{abstract}

The LIGO-Virgo-KAGRA (LVK) collaboration has detected two binary
neutron star (BNS) mergers, GW170817 and GW190425.  While no clear
electromagnetic counterpart was discovered for GW190425, a coincident
counterpart to GW170817 (GRB170817A, AT~2017gfo) was discovered in an
early-type galaxy, NGC~4993.  Using broad-band photometry and an
integrated optical spectrum, we measure a mass-weighted age of
12.4\err{0.5}{0.7}~Gyr (90\% credible) for NGC~4993, which represents
a reasonable estimate for the delay time between star formation and
merger for the GW170817 progenitor system.  BNS systems are expected
to have a power-law delay-time distribution (DTD) with a minimum delay
time, $t_{\rm min}$, of $\sim$10~Myr and a power-law index of $\alpha
\approx -1$.  This canonical DTD, when convolved with the cosmic
star-formation history, predicts $\sim$16 BNS mergers at delay times
of $\sim$100~Myr for every BNS merger at 10~Gyr; under it, the chance
of detecting even one merger as old as the GW170817 system is
$\sim$1.8\%, and 55 additional BNS merger detections are necessary for
it to be expected.  The long delay time of GW170817 therefore implies
many ``missing'' shorter-delay mergers if the canonical DTD holds.  A
DTD with $t_{\rm min} = 12.6$~Gyr and $\alpha = -1$ is favored over
the canonical DTD by a factor of 40--140, and over short gamma-ray
burst and Galactic double-NS DTDs by factors of 80--280.  These
results are in tension with current population synthesis models, the
Milky Way double NS population, and the SGRB population.  Explaining
the LVK population requires either a dramatically different separation
distribution than commonly assumed, an additional channel with a long
delay time and comparable rate to an isolated-binary formation
scenario, or the unlikely possibility of a large population of BNS
mergers missed in the LVK analysis.  This result has wide-ranging
implications for binary-population synthesis, chemical enrichment
models, connections between BNS mergers and SGRBs, the Galactic double
NS population, and predictions for future gravitational wave
observations.

\end{abstract}


\section{Introduction} \label{s:intro}

The detection of GW170817 by the LIGO-Virgo collaboration
\citep{Abbott17:gw170817} was the first unambiguous detection of a
binary neutron star (BNS) merger.  The coincident electromagnetic (EM)
counterpart \citep[GRB170817A, AT~2017gfo, SSS17a;][]{Abbott17:grb,
  Coulter17, Goldstein17} is currently the only such EM detection of a
gravitational wave (GW) event.  The LIGO-Virgo-KAGRA (LVK)
collaboration has since detected a second BNS merger, GW190425
\citep{Abbott20:0425}, however, its localization was especially poor
and no EM counterpart was detected \citep{Coughlin19:0425,
  Hosseinzadeh19, Lundquist19, Antier20, Chang21, Coulter24, Paek24,
  Smartt24}.  As a result, GW170817 represents the only GW event that
has been localized to a galaxy.

Connecting an astrophysical transient to a stellar population provides
critical information about the system that produces the transient.
For instance, the ages of stellar populations have been used to
distinguish different progenitor scenarios for various core-collapse
supernovae \citep[e.g.,][]{Schulze21, Taggart21}.  The rate of events
as a function of the stellar population age can be used to determine a
delay-time distribution (DTD); the distribution of times between star
formation and the transient event.  Such measurements have been made
for Type Ia supernovae, providing discriminating information for their
progenitor systems \citep[e.g.,][]{Maoz17}.

A power-law DTD is inevitable for any binary population where the
separation distribution is a power law,
\begin{equation}
    \frac{dN}{d\log a} \propto a^{p}
\end{equation}
\citep[where $p = 0$, a distribution that is flat in $\log a$, is
  \"Opik's Law; \citealt{Opik24}; e.g.,][]{Poveda04, Kouwenhoven07,
  Reipurth07, Sana11}, where $N$ is the cumulative number of systems
that are born\footnote{For BNS mergers, we consider the separation
distribution after both neutron stars have formed.  At this point,
gravitational radiation dominates the future evolution, and the merger
timescale is generally believed to be much longer than the stellar
evolution timescale.} at a semi-major axis smaller than $a$.  If the
inspiral is driven by gravitational radiation, the inspiral time,
$t_{\rm insp}$, is a function of the separation,
\begin{equation}
  t_{\rm insp} \propto a^{4},
\end{equation}
and the inspiral time is roughly the delay time (i.e., the time for
stellar evolution is relatively short).  In detail, the inspiral time
also depends on the eccentricity of the system and the component
masses; however, these parameters should not significantly affect the
shape of the DTD unless they evolve, on average, with cosmic time.
For the above case,
\begin{equation}
  \frac{dN}{dt} \propto \frac{dN}{da} \frac{da}{dt} \propto a^{p-1} t^{-3/4} \propto t^{p/4 - 1}.
\end{equation}
Thus for a large range of values for $p$, the power-law index,
$\alpha$, for $dN/dt \propto t^{\alpha}$ is roughly $-1$
\citep{Piran92}.  The ``classic'' formation scenario from isolated
field binaries naturally produces such a distribution with theoretical
population synthesis studies finding $-2 \lesssim \alpha \lesssim -1$
\citep[e.g.,][]{Dominik12, Belczynski18, Chruslinska18, Kruckow18,
  Vigna-Gomez18}.

Other formation paths exist, such as dynamical formation in globular,
nuclear, or young stellar clusters \citep[e.g.,][]{Grindlay06, Lee10,
  Fragione20, Santoliquido20, Ye20, Broekgaarden22}; accretion-induced
collapse of a white dwarf to a NS \citep[e.g.,][]{Ruiter19, Wang20};
and the von Zeipel--Lidov--Kozai mechanism \citep[e.g.,][]{Hamers19},
but none of these scenarios are expected to dominate the merger rate
over the isolated field binary channel \citep[e.g.,][]{Mandel22:co}.
Nevertheless, their DTDs can have both longer typical delay times and
non-power-law shapes.

Observationally, one can potentially constrain the BNS merger DTD by
examining Galactic double NS\footnote{We use `DNS' when referring to
Milky Way NS binaries, as it is the term generally used in the
literature for these systems.  We use `BNS' when discussing SGRB and
GW progenitor systems, as that is the term generally used in the
literature by those communities.  For this work, there is no physical
difference.} (DNS) systems and short gamma-ray bursts (SGRBs).  Some
DNS systems have inspiral times of less than a Hubble time, providing
a sample that should represent a subset of the present-day merging BNS
population.  SGRBs likely have binary compact-object mergers where at
least one object is a NS \citep[see][for a review]{Berger14}.
GW170817 also had a coincident SGRB \citep{Abbott17:grb, Goldstein17},
further connecting SGRBs to BNS mergers.  However, both Galactic DNS
systems and SGRBs likely have significant selection effects where the
former requires one NS to be a pulsar and the latter require specific
environments to produce a luminous afterglow.  Moreover, it is
possible that a significant fraction of SGRB progenitor systems are
not BNSs.  Therefore, one should not immediately equate Galactic DNS
systems or SGRB progenitor systems with the full merging BNS
population.

Nevertheless, recent studies have found that the Galactic DNSs and
SGRBs have power-law DTDs with $-2 \lesssim \alpha \lesssim -1$
\citep{Zevin22, Maoz24}.  In particular, \citet{Zevin22} found $\alpha
= -1.8$ with a minimum delay time of 180~Myr, which while generally
consistent with theory is a steeper distribution with a longer minimum
delay time than typically found by binary population synthesis
\citep[although see][]{Zheng07}.  \citet{Maoz24} found that the
Galactic DNS population is best described by a broken power law with
indices of $-1.1$ and $-1.9$ for the slow- and fast-merging
populations, respectively, with a best-fit $\alpha \approx -1.3$ for a
single power law fitted over the full range.

A more direct measurement of the BNS merger DTD is from GW events,
which do not suffer from significant selection effects in delay time.
However, their EM localization can.  Only GW170817 has a measured
delay time, and the discovery of the counterpart \citep{Coulter17} and
many independent detections \citep[see][]{Abbott17:mma} prioritized
massive or star-forming galaxies.  Nevertheless, AT~2017gfo was
detected by an independent, untargeted search \citep{Soares-Santos17},
mitigating this concern for the current sample.

Similar to what has been done for Type Ia supernovae \citep{Maoz12},
\citet{Safarzadeh19} suggests a method of constraining the BNS merger
DTD using detailed star-formation histories (SFHs) for BNS merger host
galaxies.  They find that 100--300 host galaxies with well measured
SFHs are necessary to reasonably constrain the DTD.  However, this
study looked at precisely constraining the full DTD instead of looking
for any constraints and does not include additional constraints from
the volumetric rate.

In this Letter, we combine the SFH for NGC~4993, the host galaxy of
GW170817, the total number of GW-detected BNS mergers, the volume-time
of the LIGO-Virgo-KAGRA (LVK) observatories, and the measured
volumetric BNS merger rate to constrain the BNS merger DTD.  In
Section~\ref{s:n4993}, we present a new measurement of the SFH of
NGC~4993 and delay time for GW170817.  In Section~\ref{s:dtd}, we use
the new GW170817 delay time along with other data to constrain the BNS
merger DTD, finding that the data disfavor the DTDs of isolated binary
evolution, SGRBs, and Galactic DNS systems relative to DTDs with long
minimum delay times.  We discuss the broad implications and conclude
in Section~\ref{s:disc}.

We adopt the $\Lambda$CDM cosmology of \citet{Behroozi19}, which is
similar to that of \citet{Planck14}, with $H_{0} =
67.8$~km~s$^{-1}$~Mpc$^{-1}$ and $\Omega_{m} = 0.307$, giving $t_{\rm
  univ} = 13.68$~Gyr.

\section{GW170817 Delay Time}\label{s:n4993}

Since GW170817 is currently the only BNS merger with a known
counterpart, and thus the only such event where a delay time can be
determined, a precise measurement is critical to any study of the DTD.

AT~2017gfo was located 2~kpc from the center of NGC~4993, an S0 galaxy
\citep{Coulter17}.  No globular cluster (GC) or star-forming region
was seen near the KN position \citep[e.g.,][]{Blanchard17, Levan17,
  Palmese17, Pan17:gw, Kilpatrick22}, and thus it is likely that the BNS
system formed both a significant time prior to merger and at a
location far from its position at the time of merger.  Thus properties
of the stellar population at the KN location are unlikely to be more
representative of the progenitor system than global properties.  We
therefore focus on modeling the global stellar population of NGC~4993.

Previous works have analyzed NGC~4993, finding typical stellar ages of
$\sim$10~Gyr (see Table~\ref{t:sfh}), all significantly older than the
typical delay time for SGRBs \citep{Nugent22}.  The galaxy has several
shells indicative of a recent ($\sim$1~Gyr ago) minor merger
\citep{Blanchard17, Levan17, Palmese17, Ebrova20, Kilpatrick22}.  A
careful treatment of the recent star formation is necessary to rule
out a relatively short delay time associated with this merger.  A
combination of broad-band photometry and a spatially integrated
optical spectrum is particularly useful to provide accurate overall
SFR and mass estimates as well as detect any recent star formation.

\subsection{Data Acquisition and Reduction}

For consistency, we use the photometry presented by
\citet{Blanchard17}.  They provide homogeneous photometry at a Kron
radius for 20 distinct wavelengths from X-ray to radio, but we focus
on the 14 ultraviolet through infrared measurements useful for
constraining stellar populations.  While one could supplement these
data with additional photometry, the measurement uncertainties are
small and the wavelength range already covers 0.2--20~$\mu$m.

We supplement these data with a spectrum obtained with the Multi Unit
Spectroscopic Explorer (MUSE) mounted on the UT4 VLT telescope
\citep{Bacon10}.  MUSE is an integral field spectrograph, and the data
cover a roughly 70\arcsec$\times$70\arcsec\ square centered on the
position of AT~2017gfo, and covering the nucleus of NGC~4993.  The
data were obtained on 18 August 2017 with a total exposure time of
2600~sec.  MUSE has a pixel scale of 0.2\arcsec\ and spectral
resolution of $R \approx 1800$--3600 (from blue to red wavelengths).
These data were previously presented by \citet{Levan17}.

We downloaded the processed MUSE spectra (using version 2.0.3 of the
ESO pipeline) from the ESO archive.  Using a custom pipeline, we
extracted a spectrum from a 10\arcsec\ radius circular aperture
centered on the center of NGC~4993.  The pipeline produces a
sky-subtracted data cube, however, our extracted spectrum contained
significant residuals around bright sky lines.  To improve the sky
subtraction, we also extracted a rectangular aperture in a blank
region in the corner of the data cube opposite the nucleus.
Subtracting this spectrum, which has a median flux very close to zero,
consistent with being clean sky, from the galaxy spectrum resulted in
smaller sky residuals without significantly altering the flux at other
wavelengths.

We also retrieved a nuclear spectrum from the 6dF Galaxy Survey
\citep{Jones04}, taken through a 6.7\arcsec\ fiber centered on the
nucleus of NGC~4993 on 2002 May 7.  It covers 3995--8430~\AA\ in the
rest frame, extending $\sim$700~\AA\ blueward of the MUSE spectrum and
including the age-sensitive features \hgamma, \hdelta, and the G band.
The 6dF spectrum is not flux calibrated, so we median-scale it to the
MUSE spectrum over their overlap and leave the continuum shape as part
of the fitting process.

\begin{figure*}
\begin{center}
\includegraphics[angle=0,width=4.76in]{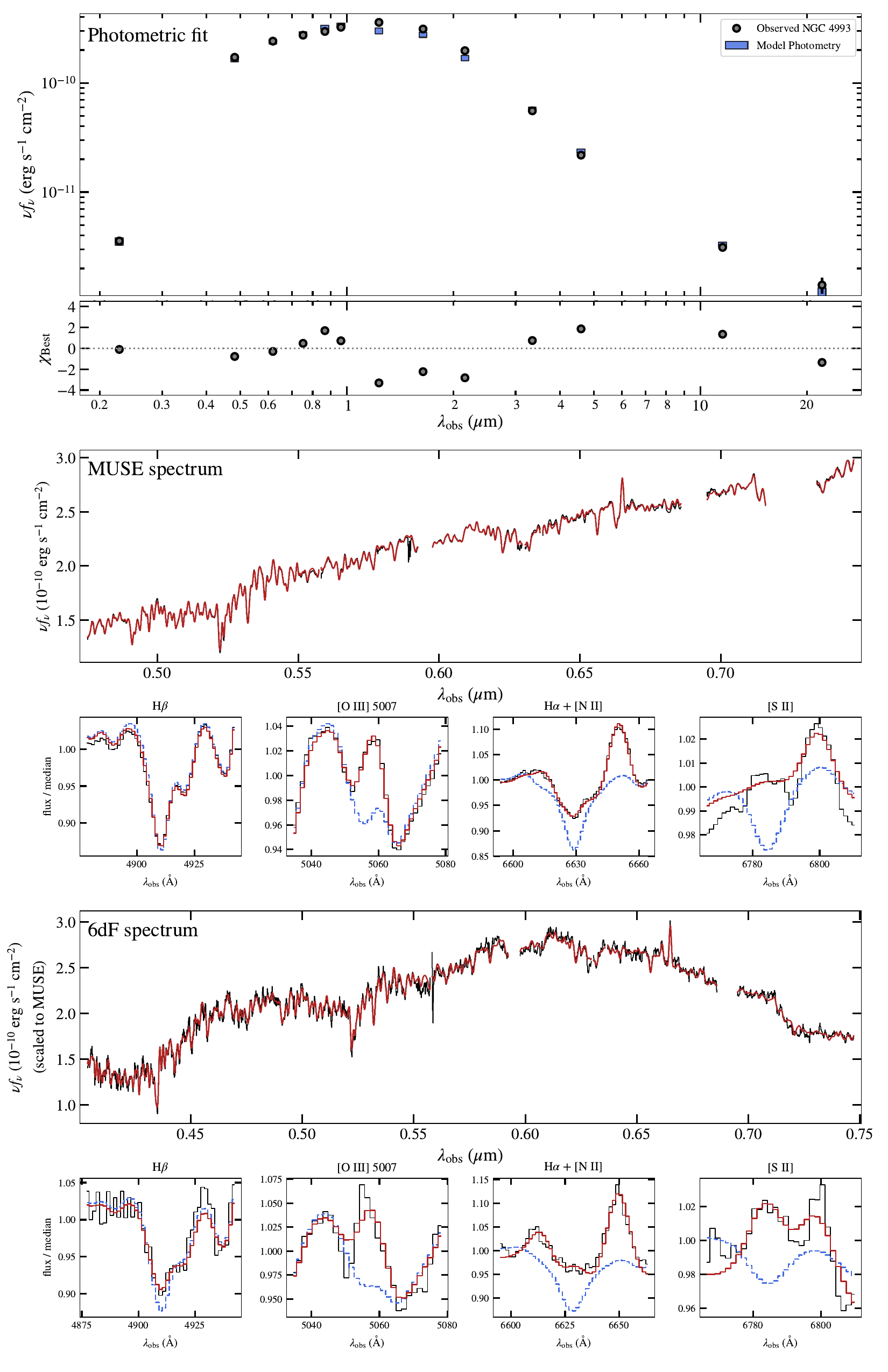}
\caption{Photometry and spectra of NGC~4993, the host galaxy of
  GW170817, with the nominal stellar-population model.  The model
  shown is the best fit of one of the four runs.  {\it Top panel}:
  Observed SED (gray circles) and the model's synthetic photometry
  through the same filters (blue boxes, whose height is the $\pm
  1\sigma$ observational uncertainty).  {\it Second panel}: $\chi =
  ({\rm model} - {\rm data})/\sigma$ for each band.  {\it Third
    panel}: MUSE spectrum (black) and the nominal model (red).  Pixels
  excluded from the fit appear as gaps.  {\it Fourth row}: MUSE
  spectrum (black) covering \hbeta, [\ion{O}{3}]~$\lambda$5007,
  \halpha+[\ion{N}{2}], and [\ion{S}{2}], each divided by its median
  flux.  The nominal model (red), which contains nebular emission,
  and the best fit without nebular emission (blue dashed) are
  overplotted.  {\it Fifth panel}: The 6dFGS nuclear spectrum, scaled
  to MUSE, with the same two models evaluated on it (not fitted to it)
  after re-optimizing only the 6dF calibration polynomial.  The
  continuum shape has not been flux calibrated.  {\it Bottom row}: The
  same as the fourth row, except displaying the 6dF
  spectrum.}\label{f:sed}
\end{center}
\end{figure*}

We present the photometry and spectra in Figure~\ref{f:sed}.

\subsection{Stellar Population Properties}\label{s:sfhfit}

We use the flexible fitting code, {\tt prospector} \citep[version
  1.3.0;][]{Leja17, Johnson21}, to fit the photometry and MUSE
spectrum of NGC~4993, where we use the python version
\citep{Johnson24} of the Flexible Stellar Population Synthesis
\citep[FSPS;][]{Conroy09, Conroy10} to create spectral-energy
distributions (SEDs) with the MIST isochrones and the MILES spectral
library.  This version of {\tt prospector} can simultaneously fit
photometry and spectra.  We used the {\tt dynesty} nested sampler
\citep{Speagle20, Koposov24} to sample the parameter space.  We use
the {\tt prospector} continuity non-parametric SFH model, with a
Student's $t$ prior on the logarithmic ratios of star-formation rate
between adjacent bins.  We assume a \citet{Chabrier03} initial mass
function.  Free parameters and their priors are $\log(M_{\star}/{\rm
  M}_{\sun})$ with a uniform prior between 9 and 12,
$\log(Z_{\star}/Z_{\sun})$ with a uniform prior between $-1.0$ and
0.19, a \citet{Kriek13} dust attenuation law with $\tau_{V}$ drawn
from a normal distribution with mean 0.3 and width 1 truncated to
0--2, a free power-law index uniform between $-1.0$ and 0.2, and a
birth-cloud component $\tau_{1} = f\,\tau_{V}$ with $f$ having a
normal prior with mean 1.0 and width 0.3 truncated between 0 and 2.

Photometric uncertainties are floored at 5\% of the flux.  The
redshift has a normal prior of $0.009783 \pm 0.0001$, the heliocentric
redshift of \citet{Hjorth17}. The spectrum, extracted from a
10\arcsec-radius aperture, is multiplied by 2.364 to match the
Kron-aperture photometry.

The spectrum's continuum is calibrated to the photometry with a
12th-order Chebyshev polynomial.  The velocity dispersion has a
uniform prior between 50 and 800~km~s$^{-1}$.  When fitting, the model
spectra are convolved with the line-spread function as a function of
wavelength \citep{Bacon17}.  There is a free spectral jitter term with
uniform prior between 1 and 10 that normalizes the spectroscopic
uncertainties relative to the photometric uncertainties.  We include a
pixel-outlier mixture model whose outlier fraction prior is uniform
between $10^{-5}$ and 0.1, with outliers given 50 times the pixel
uncertainty.  Nebular emission is included where the nebular continuum
is part of the model, the emission-line amplitudes are marginalized
analytically, and the ionization parameter is free with a uniform
prior $-4 < \log U < -1$.

We perform four independent \texttt{prospector} runs of the data with
identical settings but slightly different parameter values at the
start.  The run-to-run scatter in the recovered parameters is
comparable to the within-run uncertainty for some quantities.  All
values quoted below are therefore from the four chains pooled with
equal weight, and the quoted intervals include that scatter.

To allow for the most possible SFHs, we performed a non-parametric
fit.  A non-parametric model uses a star-formation rate and mass in
pre-defined age bins, such that the behavior between bins is not
restricted by a specific model, but the total stellar population is
matched to the data, which constrains each parameter.  This allows for
the most flexible star-formation history with multiple bursts of star
formation and an overall star-formation history that differs from
commonly used models.  A downside of a non-parametric model is that
there is significant degeneracy between two adjacent age bins if they
are sufficiently small, increasing the number of parameters without
necessarily improving the fit to the data.  Therefore, extracting
physical parameters from a non-parametric fit requires high-quality
broadband data.  This approach results in an accurate posterior
distribution of SFHs while the time resolution can capture
star-formation bursts and precisely determine the peak of star
formation.

Recent versions of {\tt prospector} allow for simultaneous photometric
and spectroscopic fitting, an option unavailable to
\citet{Blanchard17} who fit photometry and a spectrum separately,
combining posteriors; however, the general method is similar.

We fit the NGC~4993 data multiple times with different age bins.  The
oldest age bin's edge is restricted by the cosmological redshift of
NGC~4993, $z = 0.009409$, derived from the heliocentric redshift of
\citet{Hjorth17} after correction for Virgo and Great Attractor
infall.  The oldest bin ends at the age of the Universe at that
redshift, 13.68~Gyr.  We began by choosing age bins every 1~Gyr in
lookback time, which provides a relatively high-resolution
star-formation history.  Because NGC~4993 is a shell galaxy with a
possible recent merger that may have created a star-formation burst,
we added additional age bins around 1~Gyr (bin edges of 0.1, 0.25,
0.6, 0.8, 1, and 2~Gyr) to capture any significant star formation from
a potential minor galaxy merger.  We do not detect a significant
increase in star formation for this epoch.  In particular, the lack of
strong \hbeta\ absorption constrains any potential post-starburst
population.  With these age bins, most of the stellar mass formed in
the oldest bin.  To improve the temporal resolution at these early
times, which can also improve the constraints on minimum delay-time
for long-lived systems, we added 500-Myr wide bins starting at a
lookback time of 11~Gyr.  This resulted in 20 age bins and thus 20
star-formation-rate parameters.  The full set of bin edges, in Gyr of
lookback time, is 0, 0.03, 0.1, 0.25, 0.6, 0.8, 1, 2, 3, 5, 6, 7, 8,
9, 10, 11, 11.5, 12, 12.5, 13, and 13.68.  \citet{Blanchard17} present
only a magnitude limit for the \textit{GALEX} FUV measurement, and we
exclude it from the fitting.

The photometric measurements are well fit by the model, although the
overall goodness of fit, $\chi^{2} = 37.2$ for the 13 photometric
points used\footnote{We quote a $\chi^{2}$ rather than a $\chi^{2}$
per degree of freedom because the number of degrees of freedom is not
well defined here.  The non-parametric star-formation-rate parameters
are correlated through their prior, so the effective number of free
parameters is smaller than the nominal count of 20.}, indicates some
disagreement between the measurements and model.  This disagreement
appears to be caused by the three 2MASS observations, which are all
high relative to the best-fitting model.  \citet{Blanchard17} showed a
similar discrepancy when fitting the photometry with an earlier
version of \texttt{prospector}, and thus it is likely related to the
measurements rather than the model.  Since we cannot determine a
reason the 2MASS data (or any other data) would be incorrect, we
continue to include those data in the analysis.  Exclusion of the
2MASS data does not significantly affect the main conclusions of the
paper, which depend primarily on the SFR.

For the spectral fitting, we mask regions with bright sky lines and
where there is the strongest telluric absorption.  We also mask the
region around Na~D since the strength of that feature can be
influenced by Milky Way absorption at the low redshift of
NGC~4993. The MILES library ends at 7500~\AA; for redder wavelengths
FSPS uses a low-resolution theoretical library that cannot reproduce
absorption lines at the MUSE resolution, so we fit only rest
wavelengths shorter than 7400~\AA.  The spectral jitter is 1.28 and
the outlier fraction 0.007.  The FSPS models have solar-scaled
abundance ratios; the residuals at Mg~$b$ relative to the Fe features
suggest the $\alpha$-element enhancement typical of massive early-type
galaxies, which we do not model.

\begin{figure*}
\begin{center}
\includegraphics[angle=0,width=6.6in]{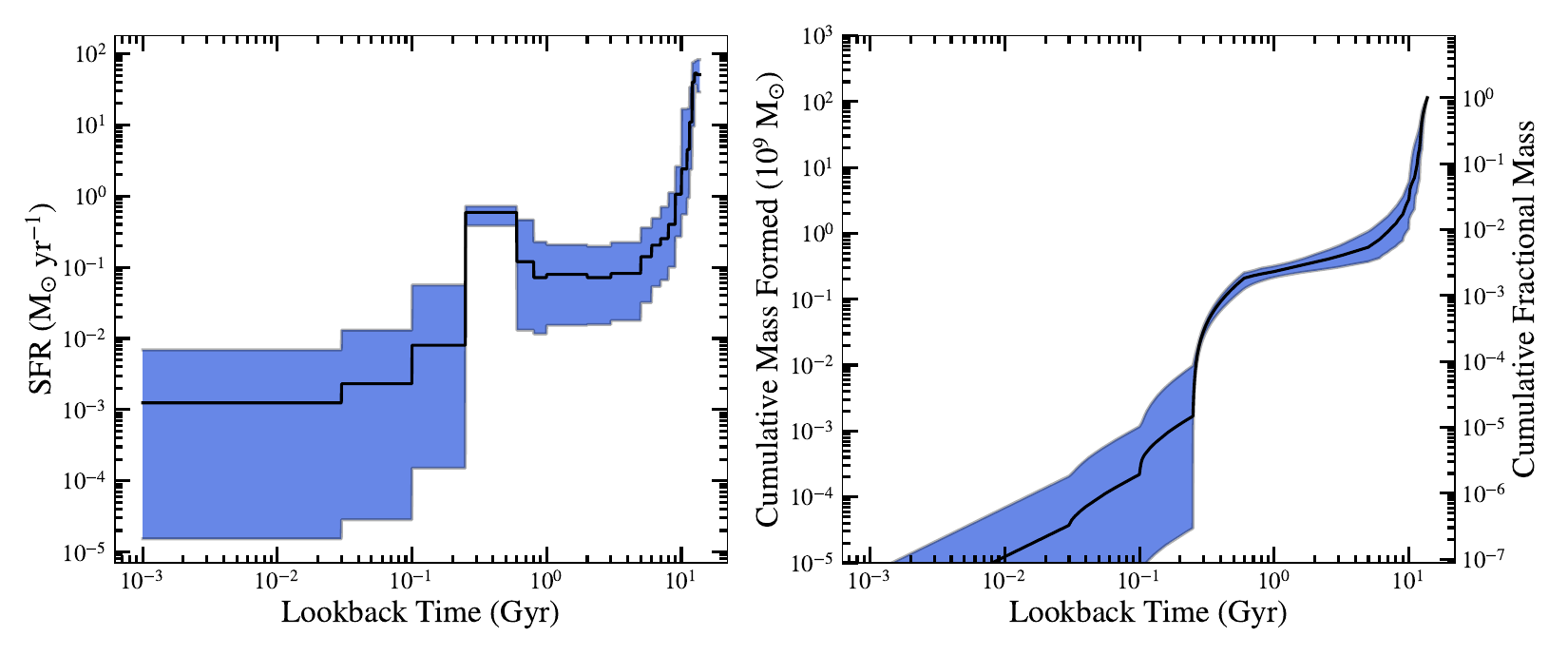}
\caption{Star-formation history of NGC~4993 (black) with the 16th to
  84th percentile range from the pooled chains in blue.  {\it Left}:
  SFR as a function of lookback time.  {\it Right}: Cumulative mass
  formed (left axis with units of $10^{9}$~M$_{\sun}$; right axis with
  units of the fraction of the median total).}\label{f:sfh}
\end{center}
\end{figure*}

\begin{figure}
\begin{center}
\includegraphics[angle=0,width=3.6in]{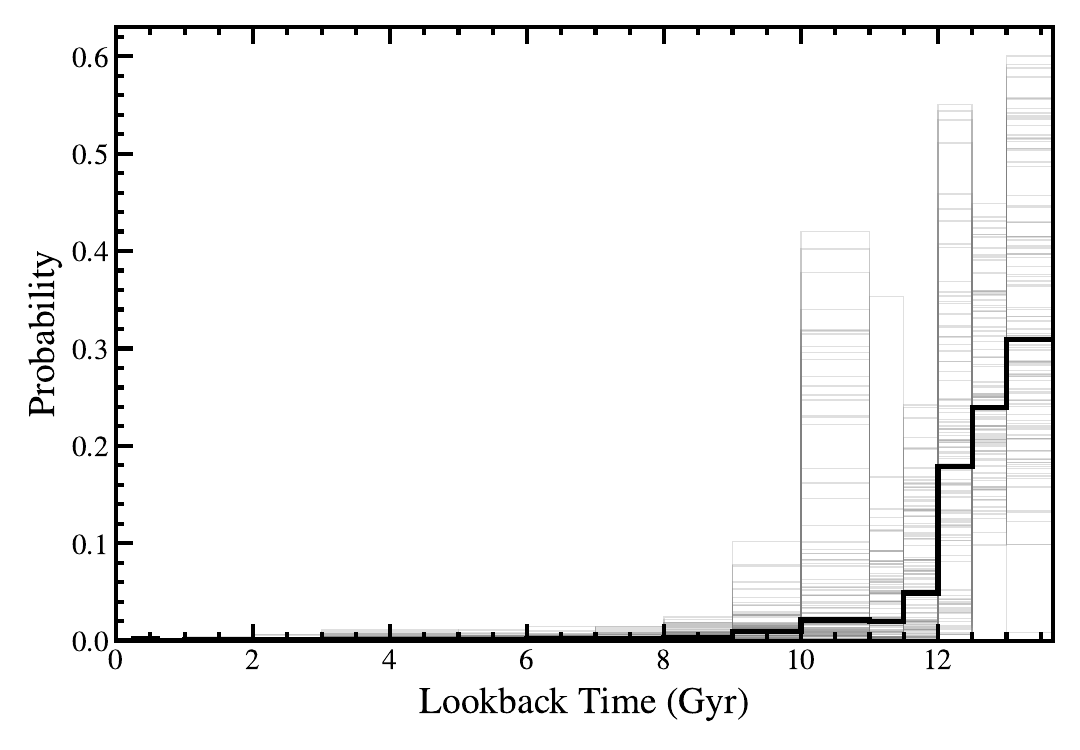}
\caption{Probability distribution for the age of stars in NGC~4993,
  corresponding to the fraction of the mass formed in each age bin.
  The grey lines are 100 random draws, taken in proportion to their
  importance weights, from the pooled four-run posterior.  The thick
  black line is the weighted median in each bin.}\label{f:pdf}
\end{center}
\end{figure}

\begin{deluxetable}{llll}
\renewcommand{\arraystretch}{1.2}
\tabletypesize{\footnotesize}
\tablewidth{0pt}
\tablecaption{Stellar Population Properties for NGC~4993\label{t:sfh}}
\tablehead{
\colhead{} &
\colhead{} &
\colhead{Mass-Weighted} &
\colhead{Half-Mass}\\[-0.3cm]
\colhead{} &
\colhead{} &
\colhead{Age} &
\colhead{Time}\\[-0.3cm]
\colhead{Reference} &
\colhead{$\log (M_{\star}/ {\rm M}_{\sun}$)} &
\colhead{(Gyr)} &
\colhead{(Gyr)}}

\startdata
\citet{Blanchard17} & 10.65\err{0.03}{0.03}  & 13.2\err{0.5}{0.9} & 11.2\err{0.7}{1.4}\\
\citet{Im17}         & 10.58                 &  6.5 & \nodata\\
\citet{Palmese17}    & 10.47\err{0.09}{0.11} & 11.3\err{0.05}{0.05} & \nodata\\
\citet{Pan17:gw}     & 10.49\err{0.08}{0.20} & 10.97 & \nodata\\[0.1cm]
\hline
This Work            & $10.83 \pm 0.01$      & 12.4\err{0.5}{0.7} & 12.6\err{0.5}{0.7}
\enddata
\tablecomments{For this work, the ages are posterior medians with 90\% credible intervals.  The mass is the present-day stellar mass of the best fit. The mass formed is $\log (M / {\rm M}_{\sun}) = 11.05$.}

\end{deluxetable}

The resulting stellar population is one with very little star
formation over the last several Gyr (Figures~\ref{f:sfh} and
\ref{f:pdf}).  The median SFH peaks at 54~M$_{\sun}$~year$^{-1}$ in
the 12.5--13~Gyr bin and is below 1~M$_{\sun}$~year$^{-1}$ for the
last 9~Gyr.  Integrating over the SFR, the posterior places
96\err{3}{6}\% (posterior median and the 5th--95th percentile
interval) of the stellar mass formed before a lookback time of
10.2~Gyr, and 50\err{35}{28}\% before 12.6~Gyr.  The mass-weighted age
of stars in NGC~4993 is 12.4\err{0.5}{0.7}~Gyr, and the logarithm of
the total mass formed (in M$_{\sun}$) is $11.05 \pm 0.02$.  After
stellar mass loss, the best fit's present-day stellar mass is
$\log(M_{\star}/{\rm M}_{\sun}) = 10.83 \pm 0.01$ (where the
uncertainty is the run-to-run scatter); we compare this value to
literature measurements in Table~\ref{t:sfh}.

We test the robustness of the results by replacing the MUSE spectrum
with the 6dF spectrum.  The wavelength range, signal-to-noise ratio,
and line-spread function differ, but all model parameters are
identical.  Both data combinations prefer a predominantly old
population. The nuclear 6dF spectrum prefers an older and more
metal-rich population with a mass-weighted age of
13.1\err{0.2}{0.6}~Gyr, 91\% of the mass formed more than 12.6~Gyr
ago, and $\log(Z_{\star}/Z_{\sun}) = -0.05$, against 12.4~Gyr, 50\%,
and $-0.11$ for the MUSE fit.  Each spectrum distinguishes the two
solutions.  Moving the 6dF fit's star-formation history and physical
parameters onto the MUSE data, with the MUSE nuisance parameters kept
at their best fit, lowers $\ln \mathcal{L}$ by 350; moving the MUSE
solution onto the 6dF data lowers it by 27.  This transfer moves the
metallicity together with the SFH, so the difference is not an age
difference alone.  The 6.7\arcsec\ fibre samples the nucleus, which is
older and more metal-rich than the light integrated over 10\arcsec.
The DTD analysis requires the star-formation history of the galaxy as
a whole, so we adopt the MUSE-spectrum-constrained SFH as nominal.
Adopting the 6dF fit instead would move more of the mass into the
oldest bins.

\section{Delay-Time Distribution Constraints}\label{s:dtd}

Assuming that BNS merger progenitor systems are created at the same
rate as the stellar mass in NGC~4993, the SFH determined in
Section~\ref{s:n4993} is the probability distribution of the age of
the GW170817 progenitor system, which ignores potential metallicity
effects.  Without an EM counterpart or particularly small GW
localization volume for GW190425, we have no prior information on the
age of its progenitor system.

For the BNS mergers, we assume a power-law DTD,
\begin{equation}
  \begin{aligned}
    {\rm DTD} &=
      \begin{cases}
          A t^{\alpha} \qquad & t_{\rm min} \le t \le t_{\rm univ}\\
          0 \qquad &{\rm otherwise},
      \end{cases}\\
  \end{aligned} \label{e:dtd}
\end{equation}
where $t$ is the delay time, $A$ is the number of BNS mergers per unit
stellar mass formed per unit delay time, and $t_{\rm univ} =
13.68$~Gyr at the redshift of NGC~4993.  We do not fit for a maximum
delay time, which cannot be constrained by the current data where the
delay time to GW170817 is similar to $t_{\rm univ}$.

Combining the DTD with the cosmic star-formation rate (CSFR),
$\psi(t)$, produces a distribution of delay times for BNS systems
merging today.  While the peak of a power-law DTD is at its minimum
delay time, there were more stars forming at $z \approx 2$ than at $z
= 0$.  Integrating this function gives the current volumetric BNS
merger rate,
\begin{equation}
  R = A \int_{t_{\rm min}}^{t_{\rm univ}} \psi(t)\, t^{\alpha} \, dt. \label{e:rate}
\end{equation}
Equivalently, the formation efficiency (i.e., the number of BNS mergers per unit stellar mass formed) is
\begin{equation}
  \epsilon = A \int_{t_{\rm min}}^{t_{\rm univ}} t^{\alpha}\, dt,
\end{equation}
 so that
\begin{equation}
  R = \epsilon \, \langle \psi \rangle_{w}, \label{e:eps}
\end{equation}
where $\langle \psi \rangle_{w}$ is the DTD-weighted mean CSFR and is a
function of $(t_{\rm min}, \alpha)$.  For a fixed $R$ and $\alpha$, as
$t_{\rm min}$ increases, so does $A$.  For a given $VT$ hypervolume,
the expected number of detections is
\begin{equation}
  \lambda = R\,VT.
\end{equation}

GW detections form an inhomogeneous Poisson process in delay time with
intensity $A\, VT\, \psi(t)\, t^{\alpha}$.  If $\mathcal{L}_{i}(t)$ is
the likelihood of the data for event $i$ as a function of its delay
time, the likelihood for $N$ detections is
\begin{equation}
  \mathcal{L} = e^{-\lambda} \prod_{i=1}^{N} \lambda\, O_{i}, \label{e:like}
\end{equation}
where
\begin{equation}
  O_{i} = \ddfrac{\int_{t_{\rm min}}^{t_{\rm univ}} \mathcal{L}_{i}(t)\, \psi(t)\, t^{\alpha}\, dt}{\int_{t_{\rm min}}^{t_{\rm univ}} \psi(t)\, t^{\alpha} \, dt}.
\end{equation}

The rate enters once per event, and $e^{-\lambda}$ is the only
counting term; writing the counting term as the Poisson probability of
$N$ detections instead changes $\mathcal{L}$ only by the constant
$N!$.  For GW170817 the delay-time information comes from its host galaxy,
\begin{equation}
  \mathcal{L}_{1}(t) \propto \frac{{\rm SFH}(t)}{\psi(t)},
\end{equation}
so that
\begin{equation}
  \lambda\, O_{1} \propto A\, VT\, \int_{t_{\rm min}}^{t_{\rm univ}} {\rm SFH}(t)\, t^{\alpha}\, dt, \label{e:host}
\end{equation}
where ${\rm SFH}(t)$ is the posterior-mean SFR of NGC~4993 in each age
bin.  The mean is exact because $O_{1}$ is linear in the SFR.  The
host SFH therefore enters only the numerator for the likelihood, and
the CSFR is only in the normalization integral in the denominator.
For GW190425, $\mathcal{L}_{2}$ is flat over $0 < t < 13.36$~Gyr, so
$O_{2}$ is a constant for every $(t_{\rm min}, \alpha)$.  Therefore
GW190425 contributes only a factor of $\lambda$, and it cannot
distinguish between DTD shapes.

We evaluate the posterior on a deterministic grid in $(t_{\rm min},
\alpha)$ and the normalization.  We use the following information: (1)
the current \textsc{FullPop-4.0} estimate of the volumetric BNS merger
rate\footnote{This rate is determined by fitting a mass function for
all GW events and integrating from 1--2.5~M$_{\sun}$.  The
50th-percentile value gives $N = R\,VT = 1.46$, consistent with the $N
= 2$ BNS mergers detected.  See also
\url{https://emfollow.docs.ligo.org/userguide/capabilities.html}.}
\citep{Abac26:gwtc4pop}, $R = 92\err{157}{69}$~Gpc$^{-3}$~year$^{-1}$,
used as a log-normal prior whose logarithmic width, 0.893, is 1.23
times the width implied by reading the LVK 90\% interval as $\pm
1.645\sigma$ (0.724), so as not to overconstrain; (2) the combined
O1--O4a $VT$ for LVK as reported by \citet{Abac25:gwtc4intro} and
adjusted to account for the full range of BNS masses\footnote{This
$VT$ is measured for a merger of two 1.4-M$_{\sun}$ NSs.  The exact
hypervolume changes for BNS mergers of differing chirp mass.  The
summed O1--O4a $VT$ for a 1.4+1.4-M$_{\sun}$ binary system is $9.00
\times 10^{-3}$~Gpc$^{3}$~year \citep{Abac25:gwtc4intro}.  For an
inspiral-dominated chirp the sensitive volume scales as
$\mathcal{M}^{5/2}$ and a flat distribution across the LVK NS mass
range of 1--2.5~M$_{\sun}$, the hypervolume is $1.768 \times$ larger
than for a 1.4+1.4-M$_{\sun}$ binary system.  We assume $VT$ does not
depend on delay time, which holds for detections at the low redshifts
of the current sample.}, corresponding to $1.59 \times
10^{-2}$~Gpc$^{3}$~year; (3) the detection of two BNS mergers, through
Equation~\ref{e:like}; (4) the CSFR as measured by \citet{Behroozi19};
(5) the SFH of NGC~4993 as measured in Section~\ref{s:n4993}; (6) a
flat prior for the delay time of GW190425; (7) a flat prior of $-10 <
\alpha < 0$; and (8) a flat prior on $t_{\rm min}$ from 0.01 to
13.45~Gyr, where the upper bound is the end of the tabulated CSFR.

\subsection{The Delay-Time Distribution Normalization}\label{ss:norm}

In addition to the minimum delay time and power-law slope, the
delay-time distribution has a normalization parameter.  One can
constrain the normalization parameter with the BNS merger rate today,
$R$, or with the formation efficiency, $\epsilon$.  The rate, $R$, is
related to $\epsilon$ through the DTD-weighted mean CSFR, $\langle
\psi \rangle_{w}$ (Equation~\ref{e:eps}).  Either parameter can be
used to normalize the DTD.

The rate has a log-normal prior set by the LVK measurement.  We can
assume a log-uniform prior for the formation efficiency, where the
maximum is limited by one half of the integrated \citet{Kroupa01}
initial mass function for M$_{\rm ZAMS} > 8$~M$_{\sun}$ and a minimum
set so there should be $\gtrsim$1 merging BNS system in NGC~4993.  The
rate log-normal prior can further be transferred to a formation
efficiency prior through the Jacobian, $dR/d\epsilon = \langle \psi
\rangle_{w}$.
  
If we normalize by the rate, the posterior likelihood factorizes into
a rate part,
\begin{equation}
  \pi e^{-R\,VT} (R\,VT)^{2},
\end{equation}
and a shape part,
\begin{equation}
  \ddfrac{\int_{t_{\rm min}}^{t_{\rm univ}} {\rm SFH}(t)\, t^{\alpha}\, dt}{\int_
{t_{\rm min}}^{t_{\rm univ}} \psi(t)\, t^{\alpha} \, dt}.
\end{equation}
The rate part is a constant for fixed $R$, and thus the rate prior,
$VT$, and the Poisson counting term do not affect the marginal
likelihood when constraining the DTD shape.  The shape part is
proportional to the fraction of today’s BNS mergers that a given DTD
draws from the stars of NGC 4993.  It diverges as $t_{\rm min}$
approaches $t_{\rm univ}$ and the denominator approaches zero, while
the numerator remains finite.  The $t_{\rm min}$ posterior therefore
piles up at the upper bound of its prior, and one cannot directly
constrain $t_{\rm min}$ with this normalization.

If one chooses to normalize by the formation efficiency, marginalizing
over $\epsilon$ multiplies the shape part by $\langle \psi
\rangle_{w}$, the CSFR cancels, yielding
\begin{equation}
  \ddfrac{\int_{t_{\rm min}}^{t_{\rm univ}} {\rm SFH}(t)\, t^{\alpha}\, dt}{\int_
{t_{\rm min}}^{t_{\rm univ}} t^{\alpha} \, dt}. \label{e:jhostit}
\end{equation}
The efficiency result is the rate result evaluated as though the CSFR
had been constant in time.  The factor $\langle \psi \rangle_{w}$ acts
as an implicit prior on the DTD shape, where it penalizes DTDs whose
progenitors formed at epochs of low cosmic SFR, which can produce the
measured rate only with a large efficiency.  With the CSFR cancelled,
Equation~\ref{e:jhostit} represents the $t^{\alpha}$-weighted mean of the
host-galaxy SFH over delays above $t_{\rm min}$. Reading the LVK rate
as a likelihood of the implied $R$ instead of a prior on $\epsilon$
(that is, without the Jacobian) reproduces the rate treatment exactly.

The different normalizations are therefore two priors on one
likelihood, not two models of the data.  We report every comparison
between named DTDs in both.  We state constraints on $t_{\rm min}$ as
likelihood-ratio limits, labeled by treatment.  Throughout, a
likelihood ratio $\Lambda$ between two DTDs is the ratio of their
likelihoods with the normalization marginalized over its prior and
nothing marginalized over $t_{\rm min}$ or $\alpha$, so it carries no
prior on either.  This allows for a direct model comparison that is
independent of any prior on $t_{\rm min}$ and $\alpha$.

\subsection{Constraints on the Minimum Delay Time and Power-Law Index}\label{ss:tmin}

Direct examination of posterior distributions can constrain the
parameters for a power-law DTD model.  For the rate normalization
treatment, the $t_{\rm min}$ posterior is the upper bound of its
prior, so we cannot accurately estimate the best-fit range for $t_{\rm
  min}$ using that treatment.  However, the efficiency normalization
has a posterior mode at $12.6 \pm 0.7$~Gyr (where the uncertainties
are the scatter across the star-formation-history runs.), at the onset
of the host's peak star formation (12.5--13~Gyr).  We find $t_{\rm
  min} > 7.5 \pm 0.2$~Gyr (90\% credible) and $t_{\rm min} > 5.6 \pm
0.2$~Gyr (95\%).

\begin{figure}
\begin{center}
\includegraphics[angle=0,width=3.6in]{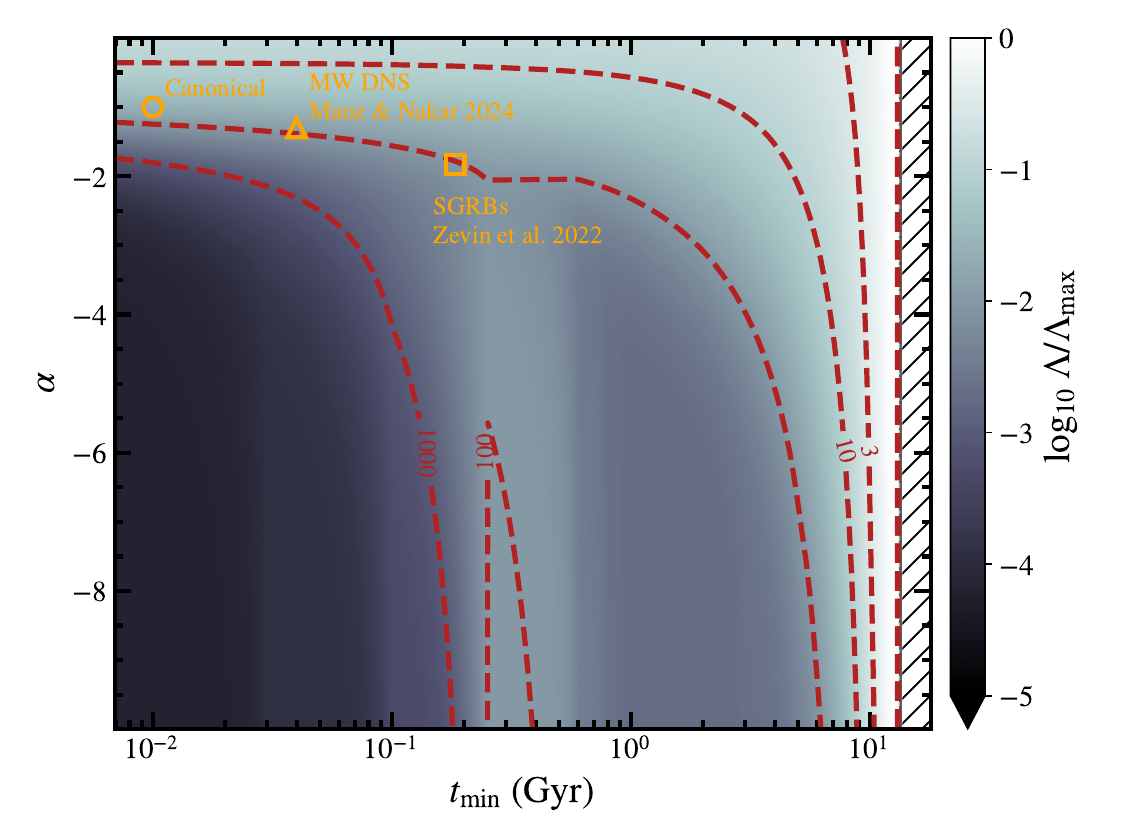}
\caption{Likelihood ratio $\Lambda(t_{\rm min}, \alpha) / \Lambda_{\rm
    max}$ across the DTD parameter plane (with the efficiency
  normalization).  For the ratio, the flat priors on both parameters
  cancel exactly.  Red curves are contours of constant likelihood
  ratio of 3, 10, 100, and 1000.  Orange symbols mark the canonical
  (circle), SGRB \citep[square;][]{Zevin22} and Galactic DNS
  \citep[triangle;][]{Maoz24} DTDs of Table~\ref{t:ratios}.  The
  hatched region represents $t_{\rm min} > t_{\rm
    univ}$.}\label{f:plane}
\end{center}
\end{figure}

Figure~\ref{f:plane} is a map of likelihood ratios between each
$(\alpha, t_{\rm min})$ model relative to the maximum likelihood model
using the efficiency treatment, $\Lambda / \Lambda_{\rm max}$.  The
likelihood ratio allows the flat priors for each parameter to cancel
exactly.  Under the efficiency treatment, relative to the maximum,
$t_{\rm min} > 9.8 \pm 0.6$, $7.0 \pm 0.5$, and $4.9 \pm 0.6$~Gyr at
likelihood ratios of 1:3, 1:10, and 1:20, respectively.  Under the
rate treatment there is no usable maximum, but we can still compare
models.  For instance, $t_{\rm min} = 10$~Myr is disfavored relative
to $t_{\rm min} = 12.6$~Gyr by $\Lambda(10~{\rm
  Myr})/\Lambda(12.6~{\rm Gyr}) = (2.0 \pm 0.4) \times 10^{-3}$, or
about 1:510.

\begin{figure*}
\begin{center}
\includegraphics[angle=0,width=6.7in]{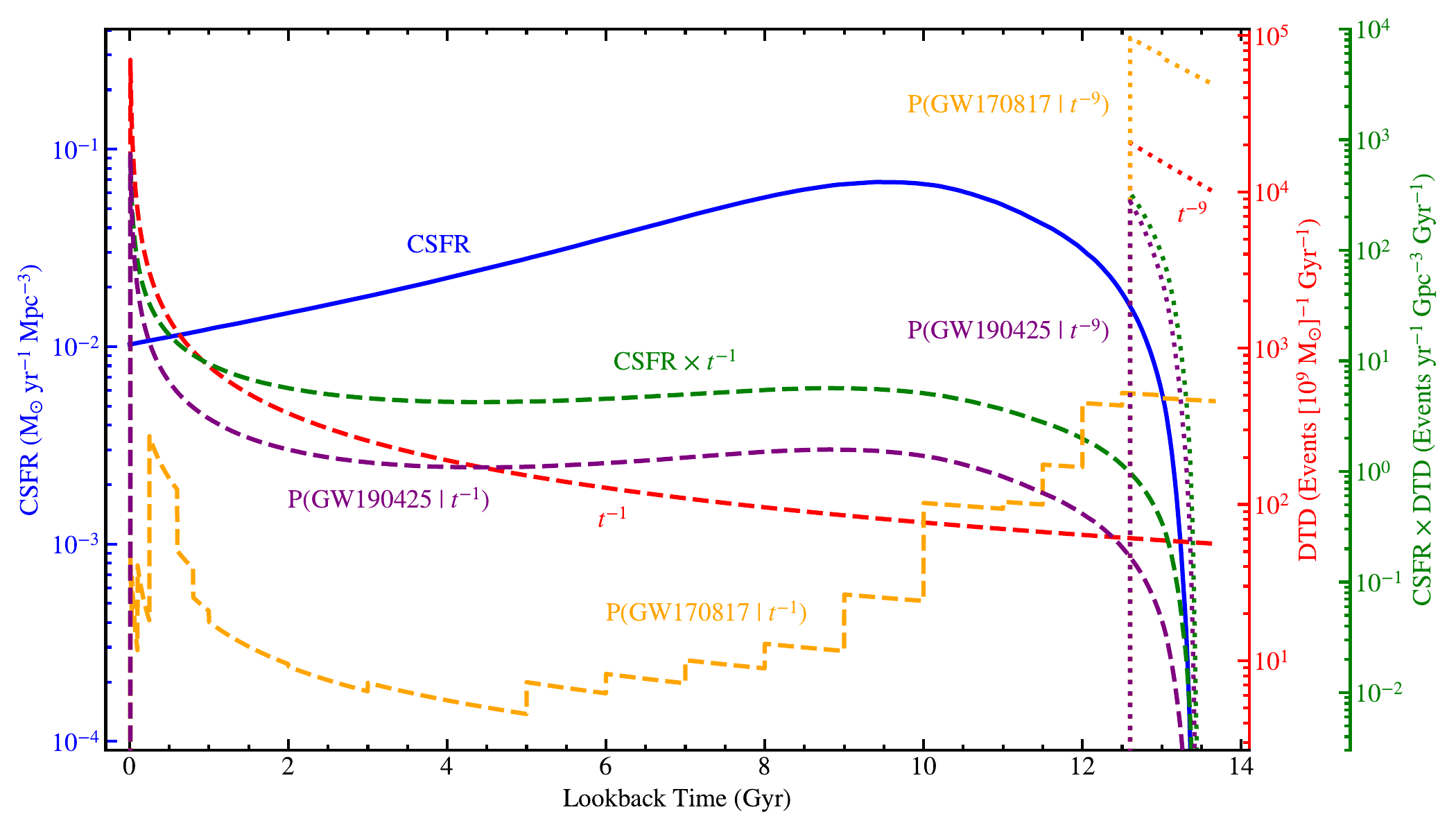}
\caption{Different components of the likelihood for two DTDs, each
  normalized to $R = 92$~Gpc$^{-3}$~year$^{-1}$.  The red-dashed line
  has $t_{\rm min} = 10$~Myr and $\alpha = -1$, and the red-dotted
  line has $t_{\rm min} = 12.6$~Gyr and $\alpha = -9$.  We also
  present the CSFR \citep[blue solid line;][]{Behroozi19}, and the
  CSFR multiplied by the DTD, which produces the distribution of delay
  times for BNS systems merging today (green lines, dashed and dotted
  to match).  The orange and purple lines are the delay-time
  probability of GW170817 (from the SFH of NGC~4993) and of GW190425
  (flat), respectively, for each DTD; the curves for the $t^{-1}$ DTD
  are multiplied by 10 for visibility, as labeled.  Their integrals,
  divided by that of the corresponding green curve, are the per-event
  factors $O_{i}$ of Equation~\ref{e:like}.}\label{f:likelihood}
\end{center}
\end{figure*}

To illustrate why long delay times are preferred, we present
Figure~\ref{f:likelihood}.  The distribution of observed BNS delay
times, which is ${\rm CSFR} \times {\rm DTD}$, is relatively flat over
$1 \lesssim t_{\rm delay} \lesssim 11$~Gyr for a $t^{-1}$ DTD, so a
canonical DTD rarely produces a merger today with a delay time as long
as that of GW170817.  A DTD with a long $t_{\rm min}$ instead draws
all of today's mergers from the oldest star formation.  NGC~4993
formed half of its stars more than 12.6~Gyr ago, whereas the Universe
as a whole formed only 1.1\% of its stars that early.  The likelihood
ratio between such a DTD and the canonical one is set mainly by the
ratio of these fractions.

\subsection{Comparison to Compact-Object Merger DTDs}\label{ss:compare}

Because the likelihood ratio between two fully specified DTDs carries
no prior on $t_{\rm min}$ or $\alpha$, it is the cleanest way to
compare possible DTDs.

\begin{deluxetable}{lccc}
\tabletypesize{\footnotesize}
\tablewidth{0pt}
\tablecaption{Likelihood Ratios for Named DTDs\label{t:ratios}}
\tablehead{
\colhead{} &
\colhead{} &
\multicolumn{2}{c}{$\Lambda / \Lambda_{\rm canonical}$}\\
\colhead{DTD} &
\colhead{$(t_{\rm min}, \alpha)$} &
\colhead{Rate} &
\colhead{Efficiency}}
\startdata
Canonical          & (10~Myr, $-1$)     & $1$ & $1$\\
\citet{Zevin22}    & (180~Myr, $-1.83$) & $0.486 \pm 0.016$ & $0.381 \pm 0.013$\\
\citet{Maoz24}     & (40~Myr, $-1.3$)   & $0.623 \pm 0.006$ & $0.529 \pm 0.005$\\
                   & (6~Gyr, $-1$)      & $2.86 \pm 0.02$ & $8.01 \pm 0.06$\\
                   & (10~Gyr, $-5$)     & $6.02 \pm 0.65$ & $15.4 \pm 1.7$\\
                   & (10~Gyr, $-1$)     & $9.66 \pm 0.06$ & $20.3 \pm 0.1$\\
                   & (12.6~Gyr, $-9$)   & $116 \pm 26$ & $40.8 \pm 9.3$\\
                   & (12.6~Gyr, $-1$)   & $138 \pm 32$ & $40.6 \pm 9.4$
\enddata
\tablecomments{Rate and Efficiency are the two treatments of the DTD normalization (Section~\ref{ss:norm}); in each, the normalization is marginalized over its prior.  The \citet{Maoz24} comparison uses their single-power-law fit over the full $\tau_{\rm gw}$ range.  \citet{Maoz24} does not fit for a minimum delay time, so we choose 40~Myr, which is consistent with their data.  Uncertainties are the scatter across the four star-formation-history runs.  The 12.6-Gyr rows have the largest run-to-run scatter: they follow the fraction of the host's mass formed more than 12.6~Gyr ago, which ranges from 0.000 to 0.95 across the SED-model variants of Section~\ref{s:disc}.}
\end{deluxetable}

Table~\ref{t:ratios} presents the likelihood ratio between various
$(t_{\rm min}, \alpha)$ combinations, including the best-fit values
for MW DNSs and SGRBs, and the canonical DTD of $(10 {\rm ~Myr}, -1)$.
Notably, the canonical DTD performs better than both of the MW DNS and
SGRB DTDs \citep{Zevin22, Maoz24}, with factors of 1.6--1.9 or
2.1--2.6, depending on normalization.  Nevertheless, all three are
disfavored relative to DTDs with long minimum delay times.  A
$(12.6~{\rm Gyr}, -1)$ DTD, where all BNS merger systems were formed
by the time half of the stellar mass of NGC~4993 had formed, is
favored over the canonical DTD by a factor of 41--138.  Similarly,
this long-delay DTD is favored over the MW DNS and SGRB DTDs by factors
of 77--280.

\section{Discussion and Conclusions}\label{s:disc}

We find that the data associated with the GW-discovered BNS merger
sample disfavor short minimum delay times relative to long ones.  For
both normalizations, a $(12.6~{\rm Gyr}, -1)$ DTD is favored over the
canonical $(10~{\rm Myr}, -1)$ DTD by a factor of 40--140, and over
the SGRB and Galactic DNS DTDs by factors of 80--280.  With the
efficiency normalization, where we can place quantifiable constraints,
$t_{\rm min} > 7.5 \pm 0.2$~Gyr (90\% credible).  The minimum delay
times favored by the data are orders of magnitude longer than any
prediction for isolated field binaries.  We do not place a marginal
constraint on the power-law index, because once $t_{\rm min}$ is large,
any power law is effectively a $\delta$ function.  However, if we
require a short delay time, the power-law index is constrained, and
steep indices are excluded for those scenarios.

A canonical DTD with $\alpha = -1$ and $t_{\rm min} = 10$~Myr produces
10 and 100 BNS mergers per unit delay time at a delay time of 1~Gyr
and 100~Myr for every BNS merger per unit delay time at a delay time
of 10~Gyr, respectively.  Convolving such a power-law DTD with the
CSFR, for every 10-Gyr merger seen today, there should be 2 and 16
mergers per unit delay time with delay times of 1~Gyr and 100~Myr,
respectively.  Under the canonical DTD, the fraction of detected
mergers with delay times as long as that of GW170817, averaged over
its delay-time distribution, is $0.018 \pm 0.005$.

\citet{Safarzadeh19} suggested that $>$100 localized events would be
necessary to constrain the DTD.  With the small, current sample, we
can answer a narrower question: are the data consistent with the
canonical DTD?  If the canonical DTD were correct, a median of 2
additional events (1--7 for 68\% of realizations) would be required to
be consistent at 90\% credibility and 55 additional mergers are
necessary for a GW170817-like delay time to be expected.  For a
canonical DTD to be statistically consistent with the data, there must
be ``missing'' mergers at short delay times.

The exceptional nature of GW170817's environment, relative to
expectations, is particularly informative.  With only one localized
event and only two total events, we have placed strong and surprising
limits on the minimum delay time for BNS mergers.  The lack of several
detected BNS mergers with shorter delay times than GW170817 is in
tension with the generally accepted progenitor pathway for BNS
mergers.  These ``missing'' mergers require re-examination of the
assumptions for these systems.

The ``missing mergers'' problem can be solved by either (1) a
dramatically different initial separation distribution for BNS systems
than typically assumed, (2) an additional, perhaps dominant, formation
channel with a long delay time, or (3) a large number of BNS mergers
missed by LVK.

We find it extremely unlikely that any mergers would be missed by the
LVK analysis.  Besides the large number of other compact-object
mergers discovered by LVK, independent analyses also recover the same
BNS events, even when considering atypical waveforms \citep{Nitz23}.

We also believe it is unlikely that the separation distributions would
be significantly different from those assumed.  Such a distribution
would be significantly different from that of the Galactic DNS systems
\citep[see][and references therein]{Vigna-Gomez18} and from massive
binary stars \citep{Sana12}.  A minimum delay time of 10~Gyr would
correspond to a minimum initial separation of 0.021~AU for two
1.5~M$_{\sun}$ objects in a circular orbit.

It seems inevitable that at least some BNS systems that merge within a
Hubble time should form from isolated binary evolution.  For instance,
there is a Galactic DNS system with an inspiral time of only 85~Myr
\citep{Burgay03}.  This BNS population has a power-law DTD with a
short minimum delay time and power-law index of $-2 \lesssim \alpha
\lesssim -1$ \citep{Maoz24}. Assuming the above identified issues are
not the cause of the ``missing'' mergers, then the only viable
solution is another formation channel with a long minimum delay time
that has a comparable or even dominant rate to the field formation
channel.

Perhaps the most promising alternative pathway is dynamical formation
in dense stellar systems \citep{Fabian75}, which can have a long
minimum delay time.  For GCs, the delay times are driven by the
cluster core-collapse time and period of black hole burning, which is
generally similar to a Hubble time \citep{Hopman06, Kremer20}.  This
progenitor channel is natural for GW170817 \citep{Palmese17}, despite
the fact that there is no detected GC at or near the location of
AT~2017gfo to deep limits \citep{Kilpatrick22}.

One confirmed and three unconfirmed Milky Way DNS systems \citep[of
  $\sim$24 discovered, e.g.;][and references therein]{Tauris17,
  ColomiBernadich23, Deng24} reside in GCs \citep{Anderson90,
  Prince91, Jacoby06, Lynch12, Balakrishnan23, Padmanabh24},
demonstrating that some BNS systems likely form dynamically in GCs.
Because already merged systems (i.e., those who have short delay
times) are not detectable, it is difficult to translate the observed
population into a delay time.  Nevertheless, \citet{Maoz24} found that
the DTD for the Milky Way DNS population not currently in GCs is best
fit with a broken power law, indicating two formation channels in the
field.  Perhaps a subset of the field population originated in GCs and
were ejected.

A GC origin may also solve the ``eccentricity gap'' problem
\citep{Andrews19}, where the Milky Way DNS sample that will merge
within a Hubble time lacks systems with $0.3 \lesssim e \lesssim 0.6$.
This gap is unexpected if DNS systems are produced from a single
field-binary formation channel; however, isolated binaries may produce
low-eccentricity systems while dynamically formed systems would
naturally have higher eccentricity.  The DNS system B2127+11C in M15
\citep{Anderson90} has properties similar to the field systems with
high eccentricity, providing a loose connection between the
high-eccentricity DNS systems and a GC origin.

Although some studies have suggested that the GC BNS merger rate
should be similar to that of field binaries \citep{Grindlay06, Lee10},
recent theoretical studies have suggested the GC rate to be orders of
magnitude smaller \citep[e.g.,][]{Bae14, Belczynski18, Ye20}.  A GC
BNS merger rate higher than what recent studies predict is likely
necessary for this channel to be responsible for the long-delay BNS
mergers such as GW170817.

The DTDs favored by our data are preferred over those of the SGRB
\citep{Zevin22} and Galactic DNS \citep{Maoz24} populations by
likelihood ratios of 20--53 and 16--38, respectively, for a $(10~{\rm
  Gyr}, -1)$ DTD, and by more for $t_{\rm min} = 12.6$~Gyr.  This
tension suggests that LVK-detected BNS merging systems do not have the
same parent population as either those of SGRBs or Galactic DNSs,
unless there are strong selection effects for the samples.  SGRBs
require a strong gamma-ray signal and a luminous afterglow to be
detected and localized to a host galaxy.  The former depends on the
detailed physics of the merger while the latter requires a favorable
circumburst/stellar density.  The cosmological sample of SGRBs may not
be representative of the SGRBs associated with typical BNS mergers (if
they all have SGRBs).  Moreover, some SGRBs may be the result of
NS-black hole mergers, which would obviously be a different population
from the BNS merging systems.  Galactic DNSs have significant
selection effects where at least one NS must be a detectable pulsar
and the system can not have merged already.  While one can attempt to
debias the sample using survivor statistics, this requires several
assumptions.  Finally, GW190425 has a total system mass significantly
larger than any known Milky Way DNS system \citep{Abbott20:0425} and
likely a high mass ratio \citep{Foley20:0425}, suggesting a different
progenitor pathway than most BNS systems \citep{Romero-Shaw20,
  Safarzadeh20} and further indicating that the LVK BNS and Milky Way
DNS systems are drawn from different parent populations.  We caution
future studies from equating these three observed populations.

An outcome of at least some BNS mergers is a kilonova with significant
amounts of $r$-process material \citep[e.g.,][]{Kasen17}.  BNS mergers
are expected to have a significant impact on $r$-process enrichment in
the Galaxy \citep{Lattimer74, Symbalisty82, Freiburghaus99}.  Tracking
the abundances of stars, \citet{Cote19} found that a single power-law
DTD with $\alpha = -1$ could not reproduce the Eu enrichment in the
Galaxy.  Steeper power laws or adding a ``burst'' component improved
the fit, similar to the broken power law of \citet{Maoz24}.  The DTD
presented here makes this problem significantly worse since there
would be very little $r$-process enrichment from BNS mergers until
several Gyrs after star formation.  Instead, chemical enrichment
studies require an additional $r$-process component, perhaps from
field BNSs, that are distinct from the full population observed by
LVK.  A consequence is that if BNS mergers are responsible for a
significant fraction of the $r$-process elemental production at $t <
5$~Gyr, there must be a second, fast-merging channel with a merger
rate comparable to that of GW170817-like systems.  Regardless,
previous assumptions about the BNS DTD for chemical enrichment studies
must be re-evaluated.

Future EM searches for GW counterparts may wish to adjust their
strategies based on our findings.  Most galaxy-targeted searches
prioritize galaxies, in part, by their star-formation rate \citep[or a
  proxy such as $B$-band luminosity; e.g.,][]{Coulter24} to match an
anticipated $t^{-1}$ DTD.  Since the measured DTD significantly
deviates from this nominal form, such a weighting may be suboptimal
for finding counterparts.

Our results will significantly affect predictions for future GW
detectors and observing runs.  While the measured BNS merger rate in
the local universe is independent of the DTD, the evolution of the
merger rate with redshift will strongly depend on the DTD shape.  If
the minimum delay time for GW170817-like events is 10~Gyr, such events
would not have enough time to merge until $z \approx 0.33$, where the
Universe is 10~Gyr old.  Our measurement of a long minimum delay time
(or the possibility of more than one progenitor channel) will likely
result in significantly lower rates at increasing redshifts than
previously predicted.

Finally, we are anxiously awaiting more GW detections of BNS mergers.
While additional localized events will greatly improve our DTD
constraints, even those without EM counterparts, similar to GW190425,
are useful for constraining the overall rate.  Observational evidence
beyond the GW data suggest multiple formation channels contributing at
comparable rates, and therefore the true DTD is unlikely to be as
simple as that assumed here (a power law with a minimum delay time).
We do not currently have enough data to constrain more complex DTDs,
and more localized events will be necessary to do so.

\begin{acknowledgments}
I extend my deepest appreciation to D.\ Maoz for his elegant
descriptions of DTDs and his exceptional work to constrain them, both
of which inspired this study.  I would like to thank J.\ Andrews,
R.\ O'Shaughnessy, P.\ Macias, E.\ Ramirez-Ruiz, S.\ Sigurdsson,
L.\ Singer, C.\ Ye, and M.\ Zevin for discussions as this project
evolved.  I would like to thank P.\ Behroozi for providing the CSFR
data files from UniverseMachine \citep{Behroozi23}.

The UCSC transients team is supported in part by NSF grant
AST--2307710 and by a fellowship from the David and Lucile Packard
Foundation to R.J.F.

This work was performed primarily at the Aspen Center for Physics,
which is supported by National Science Foundation grant PHY--2210452.

Based on data obtained from the ESO Science Archive Facility with DOI:
\url{https://doi.org/10.18727/archive/41}.
\end{acknowledgments}

\facility{VLT:Yepun(MUSE), AAT:6dF}

\bibliography{astro_refs}{}
\bibliographystyle{aasjournal}

\end{document}